# Magneto-resistance up to 60 Tesla in Topological Insulator $Bi_2Te_3$ Thin Films

S. X. Zhang[1a)], R. D. McDonald[2], A. Shekhter[2], Z. X. Bi[1], Y. Li[2], Q. X. Jia[1] and S. T. Picraux[1b)]

1. *Center for Integrated Nanotechnologies, Los Alamos National Laboratory, Los Alamos NM 87545*
2. *National High Magnetic Field Laboratory, Los Alamos National Laboratory, Los Alamos NM 87545*

Abstract

We report magneto-transport studies of topological insulator $Bi_2Te_3$ thin films grown by pulsed laser deposition. A non-saturating linear-like magneto-resistance (MR) is observed at low temperatures in the magnetic field range from a few Tesla up to *60 Tesla*. We demonstrate that the strong linear-like MR at high field can be well understood as the weak antilocalization phenomena described by Hikami-Larkin-Nagaoka theory. Our analysis suggests that in our system, a topological insulator, the elastic scattering time can be longer than the spin-orbit scattering time. We briefly discuss our results in the context of Dirac Fermion physics and 'quantum linear magnetoresistance'.

a) Present Address: Department of Physics, Indiana University, Bloomington, IN, 47405.
   Electronic mail: sxzhang@indiana.edu

b) Electronic mail: picraux@lanl.gov

The magneto-resistance (MR) of metals can be strongly affected by the details of the Fermi surface geometry and character of electron-electron (e-e) interactions[1], and therefore gives valuable insight into the physics dominating the conductivity. Furthermore, materials with nontrivial MR can provide great opportunities in magnetic sensor and memory applications. In conventional metals, the Lorenz force induced by a magnetic field bends the trajectory of an electron, affecting its motion under the action of electric field and gives rise to an increase in electrical resistance. The vector nature of magnetic field typically prevents a strong magneto-resistance response when the physics is expected to be smooth (analytic) in field, leading to a quadratic field-dependence in the low field range and saturating resistance at high fields[1]. However, in some special systems, the resistance can grow linearly with increasing magnetic field. For example, non-saturating linear MR has been observed in polycrystalline materials with open Fermi surfaces[2], in Dirac materials[3-5] where all the Dirac Fermions can be collapsed into the lowest Landau level by a strong magnetic field[6], and in inhomogeneous materials[7-9] where the mobility fluctuations play an important role[10].

Recently linear MR (up to 15 T or less) has also been observed in a new class of quantum materials, i.e. 3-d topological insulators (TIs), including $Bi_2Se_3$ thin films[11, 12] and nanoribbons[13], and $Bi_2Te_3$ single crystals[14]. TIs have unique surface states that ride on top of a single Dirac cone, and are robust against nonmagnetic impurities or disorder owning to the strong spin-orbit coupling and time-reversal symmetry[15-17]. Because of the existence of Dirac surface states, the origin of linear MR in TIs has been suggested to be associated with the Landau level splitting of the Dirac cone in the presence of a strong magnetic field[12]. However, further gated measurements manipulating charge carrier density have shown that the linear MR behavior becomes weak when the bulk conduction is suppressed, indicating that the MR may be due to bulk channel rather than the surface states[11]. Here, we report high field magneto-transport studies of the topological insulator $Bi_2Te_3$ thin films grown by pulsed laser deposition. The resistance at low temperatures shows a strong linear-like dependence on the magnetic field from a few Tesla up to *60 Tesla*, which continuously evolves from a sharp dip at low field. Such low field behavior is recognized as a signature of weak anti-localization effects. We demonstrate that the high field linear-like MR of our sample can indeed be well understood in the framework of the Hikami-Larkin-Nagaoka theory up to the highest fields measured (*60 T*).

The $Bi_2Te_3$ thin films of ~55 nm thickness were grown on $SrTiO_3$ (111) substrates by pulsed laser deposition (PLD) using a commercial stoichiometric $Bi_2Te_3$ target with a purity of 99.999%. The XeCl laser energy was ~ 0.9 J/cm$^2$, and a laser repetition rate of 0.2 Hz was used to promote high quality films[18]. The PLD chamber has a base pressure of ~$1\times10^{-6}$ Torr. Argon gas was introduced into the chamber during the growth and the pressure was kept at 300 mTorr. The substrate temperature was held at 250 $^o$C. The film growth rate was: 2.2 nm/min. X-ray diffraction (XRD) was performed using a Panalytical MRD PRO x-ray diffractometer with Cu K$\alpha$ ($\lambda$ = 1.5418 Å) radiation. The spatial distribution of the elements Bi and Te in the film was studied by energy dispersive x-ray spectroscopy (EDX) mapping in a scanning electron microcopy model (FEI Quanta 400 F). The high field Hall effect and magnetoresistance measurements were carried out at low temperatures (1.5K and 30 K) using a 60 T pulsed magnetic field (8 ms rise time) at the National High Magnetic Field Laboratory at Los Alamos National Laboratory. The magneto-resistance and Hall data were measured using standard DC measurement and a Hall bar geometry. Both field and current were reversed to symmetrize the data and eliminate parasitic voltages induced by the fast changing rate of magnetic field. The low field measurements (T<9 Tesla) were performed using a low frequency lock-in technique in a Quantum Design Physical Property Measurement System (PPMS).

The out of plane (θ–2θ) x-ray diffraction pattern (Figure 1(a)) only shows peaks corresponding to the (000n) planes of the film in addition to the substrate SrTiO$_3$ (111) plane, suggesting the films are grown along the c axis without forming any impurity phase. The rocking curve of the (0006) peak (inset of Figure 1 (a)) has a full width at half maximum ~0.22°, which is close to the best values obtained in molecular beam epitaxy grown films[19] and indicates that the films are highly oriented along the c-axis. The spatial distribution of the elements Bi and Te were shown in the EDX elemental mapping image (Figure 1 (b)) to be laterally uniform within the resolution of the probing beam, and thus to rule out the existence of any significant metal clustering.

The Hall effect and magnetoresistance measurements were performed on the thin film samples using a pulsed magnet with field strengths up to 60 Tesla at two different temperatures: 1.5 K and 30 K. The Hall resistance shows a linear dependence on the magnetic field (Figure 2 (a)) and from the slope we can determine that the majority charge carriers are electrons with a concentration of $n_e^{3D} \sim 2 \times 10^{19} cm^{-3}$. The magnetoreistance $MR_{xx} = \frac{R_{xx}(B) - R_{xx}(0)}{R_{xx}(0)}$ as a function of magnetic field is shown in Figure 2 (b). The magnetic field is perpendicular to the film surface. At both temperatures, sharp dips were observed at low fields, suggesting the weak anti-location effect (WAL)[20-27]. More low-field MR data (inset of Figure 2 (b)) were taken at different temperatures using a PPMS system, revealing the existence of WAL. Meanwhile, a non-saturating linear-like MR was observed at higher fields from a few Tesla to 60 Tesla. This high field linear-like MR is the major finding of this work and we will focus on its origin in the following paragraphs.

Previously, linear MR up to 15 T for thin films of topological insulator $Bi_2Se_3$ [12] was attributed to Abrikosov's 'quantum linear magnetoresistance' mechanism[6]. For the Abrikosov mechanism to apply however, the field must exceed the quantum limit, whereby all of the Dirac Fermions are quantized to the lowest Landau level. The magnetic field that is required to reach this quantum limit must be greater than $B^*$ where $B^* = \frac{1}{2e\hbar v_F^2} \times (k_B T + E_F)^2$, $v_F$ is the Fermi velocity and $E_F$ is the Fermi energy[6]. For our $Bi_2Te_3$ films, we estimated the Fermi level to be ~0.03 eV above the bottom of the bulk conduction band (BCB) based on the electron concentration of $n_e^{3D} \sim 2 \times 10^{19} cm^{-3}$. From the band structure (a schematic is shown in the inset of Figure 2 (a)) that was determined by angle-resolved photoemission spectroscopy (ARPES) measurements [28], the bottom of BCB is ~0.295 eV above the Dirac point, so the Fermi energy is $E_F \sim 0.325 eV$ with respect to the Dirac point. According to the ARPES[28] and quantum oscillation measurements[14], the Fermi velocity of the surface Dirac Fermions in $Bi_2Te_3$ is ~ $4 \times 10^5 m/s$. Therefore, in order to satisfy the quantum limit, $B^*$ should be ~500 T, which is much higher than the on-set field of the linear behavior observed here (a few Tesla as seen in Figure 2 (b)).

A similar discrepancy was found in another well-known linear MR system, i.e. bulk silver chalcogenides[6,7]. To address this, Abrikosov proposed a quantum model with the assumption of material inhomogeniety[6], i.e. regions with higher electron concentration are imbedded into regions with a much smaller electron concentration where the extremal quantum situation can take place. For the silver chalcogenides, this sample inhomogeniety has been evidenced by the existence of Ag clusters (size~500 nm) at the boundaries of polycrystalline grains[29]. Our EDX elemental mapping (Figure 1 (b)), however, rules out the possibility of metal clusters in our thin films. Moreover, if this inhomogeneity model applies to our system, according to the on-set field of a few Tesla, the $E_F$ for the regions where quantum limit occurs should be only ~ 0.025 eV above the Dirac point and ~ 0.105 eV below the top of the bulk valence band (red line in the upper inset of Fig. 3 (b)), and as a result, holes should be the majority charge carriers in these regions. The coexistence of multi-conduction channels of holes and electrons could lead to non-linearity in the Hall resistance curve[14, 30], which is in contrast to the observed n-type linear behavior of Figure 2 (a).

The coexistence of a sharp dip at low fields and linear-like MR at high fields suggests the possibility that both are associated with the weak antilocalization (WAL) effect. The WAL related low field MR (or magneto-conductivity) in topological insulators[20-27] has been described by the simplified Hikami-Larkin-Nagaoka (HLN) equation [31]

$$\Delta\sigma(B) = -\frac{\alpha e^2}{\pi h}[\ln(\frac{B_\Phi}{B}) - \psi(\frac{1}{2} + \frac{B_\Phi}{B})] = -\frac{\alpha e^2}{\pi h}\eta(\frac{B_\Phi}{B}) \tag{1}$$

Where $\psi$ is the digamma function, $\eta$ function is defined as: $\eta(x) = \ln(x) - \psi(\frac{1}{2} + x)$, the characteristic magnetic field $B_\Phi = \frac{h}{8\pi e D \tau_\Phi}$, $\tau_\Phi$ is phase coherence time, $D$ is the diffusion constant, $h$ is the Planck's constant and the prefactor $\alpha$ is equal to -1/2 for single coherent channel. It should be noted that this simplified equation is only valid when the following condition is satisfied: $\frac{1}{\tau_\Phi}, B \ll \frac{1}{\tau_e}, \frac{1}{\tau_{SO}}$ ($\tau_{SO}$ is the spin-orbit scattering time). When multiple parallel conduction channels (i.e. two surfaces and the bulk) are considered, the characteristic field is related to the coherence both on the surface and in bulk, and the prefactor $\alpha$ is -0.5 ~ -1.5 depending on the decoupling between channels[24, 25]. We fitted our magnetoconductivity data at 1.5 K and 30 K using equation 1. As seen in Figure 3, the fitting (green) shows a deviation from the experimental data (red) in both low field (inset of Fig. 3) and high field regions. This deviation is likely to be because the condition $B \ll \frac{1}{\tau_e}, \frac{1}{\tau_{SO}}$ is not satisfied in the entire magnetic field region. In this case, we have to apply the original HLN equation[31]:

$$\Delta\sigma(B) = -\frac{\alpha e^2}{\pi h}[2\eta(\frac{B_e}{B}) - 2\eta(\frac{B_1}{B}) + \eta(\frac{B_2}{B}) - \eta(\frac{B_3}{B})]$$

Since there is no magnetic scattering in our samples, the above equation becomes

$$\Delta\sigma(B) = -\frac{\alpha e^2}{\pi h}[2\eta(\frac{B_e}{B}) - 2\eta(\frac{B_{SO}^Z + 2B_{SO}^x + B_\Phi}{B}) + \eta(\frac{B_\Phi}{B}) - \eta(\frac{4B_{SO}^x + B_\Phi}{B})] \quad (2)$$

We found that to obtain a better fit to the experimental data, $B_e < B_{SO}$ has to be satisfied, in other words, the elastic scattering time $\tau_e$ should be longer than the spin-orbit scattering time $\tau_{SO}$. The blue curves in Figure 3 are the fittings with the following parameters: $\alpha(1.5K) = -(0.711 \pm 0.005)$, $B_e(1.5K) = (7.9 \pm 0.4)T$ and $B_\Phi(1.5K) = (0.0052 \pm 0.0003)T$ and $\alpha(30K) = -(0.88 \pm 0.05)$, $B_e(30K) = (8 \pm 1)T$, and $B_\Phi(30K) = (0.33 \pm 0.03)T$. As shown in Figure 3 and its insets, the fitting using equation (2) is better than using equation (1). For self-consistency, the $B_e$ obtained above is indeed less than the maximum magnetic field that was applied, as a result the simplified HLN equation (eq. 1) is not adequate to describe the WLA effect. Although in conventional systems the spin-orbit scattering time $\tau_{SO}$ is usually longer than the elastic scattering time $\tau_e$[32], given the strong spin-orbit coupling in topological insulators, it is possible to have $\tau_{SO} < \tau_e$. This can be understood as the following: when the spin-orbit coupling is weak as in conventional systems, the precession of the spin is slow and the corresponding spin-orbit relaxation time is longer than the elastic scattering time; While in topological insulators where spin-orbit interaction is very strong, spin-orbit field can be larger than the elastic scattering rate[33], so the spin precession is fast and spin relaxation time can be even shorter than the elastic scattering time. We note that Assaf et al.[34] recently reported linear MR up to 14 Tesla in $Bi_2Te_2Se$ thin films at higher temperatures and they suggested that it arises from the competition between a logarithmic phase coherence component and a classical quadratic component. We show that the linear-like MR observed here can be well described by the HLN theory without the consideration of classical quadratic contribution. We also note that Gao et al.[35] recently reported the emergence of linear-like MR together with the enhancement of WAL in thin $Bi_2Se_3$ sheets upon tuning the carrier density using a back gate. This further supports our conclusion that the linear MR is associated with the WAL effect. We note that strong positive MR has also been reported in some transition metal oxides, for example in $VO_x$ thin films[36], where the interaction between d-orbit electrons is strong and non-trivial. In that case, the interplay between interaction and disorder gives rise to a strong MR that is typically proportional to $H^{1/2}$ [32, 37]. However, as we have shown above, in our $Bi_2Te_3$ thin films with p-oribit conduction electrons, the linear-like MR can be described well by a quasi-particle model (i.e. the HLN theory) without considering strong/non-trivial e-e interactions. Moreover, band structure calculations without considering e-e interactions agrees well with the experimental results[38], which suggests that the e-e interaction in $Bi_2Te_3$ is indeed not as strong as in strongly correlated metal oxides.

In summary, we observed non-saturating linear-like magnetoreisistance in topological insulator $Bi_2Te_3$ thin films from a few T up to 60 T at low temperatures. Due to the larger Fermi surface, the magnetic field is not high enough to reach Abrikosov's quantum limit where all the surface Dirac Fermions are quantized into the lowest Landau level. We demonstrate that the strong linear-like MR at high fields can be well understood as the weak antilocalization phenomena described by Hikami-Larkin-Nagaoka theory. Our analysis suggests that in our topological insulator, the elastic scattering time is longer than the spin-orbit scattering time. The non-saturating linear-like MR observed here up to 60 T suggests many potential applications for topological insulators, including high-field magnetic sensor technology.

We thank Professors Alexander Finkelstein, Lincoln J. Lauhon, Babak Seradjeh and Gerardo Ortiz for helpful discussions. This work was performed, in part, at CINT, a U.S. Department of Energy, Office of Basic Energy Sciences user facility and in part at the National High Magnetic Field Laboratory, a US National Science Foundation supported Center through Cooperative Grant No. DMR901624. The research was funded in part by the Laboratory Directed Research and Development Program at LANL, an affirmative action equal opportunity employer operated by Los Alamos National Security, LLC, for the National Nuclear Security Administration of the U.S. Department of Energy under contract DE-AC52-06NA25396.

**References**

1. A. A. Abrikosov, Fundamentals of the Theory of Metals, Amsterdam, Netherlands : North Hollard (1988).
2. P. Kapitza, P R Soc Lond a-Conta **123**, 292 (1929).
3. K. K. Huynh, Y. Tanabe and K. Tanigaki, Phys Rev Lett **106**, 217004 (2011).
4. K. F. Wang, D. Graf, L. M. Wang, H. C. Lei, S. W. Tozer and C. Petrovic, Phys Rev B **85**, 041101 (2012).
5. A. L. Friedman, J. L. Tedesco, P. M. Campbell, J. C. Culbertson, E. Aifer, F. K. Perkins, R. L. Myers-Ward, J. K. Hite, C. R. Eddy, G. G. Jernigan and D. K. Gaskill, Nano Lett **10**, 3962 (2010).
6. A. A. Abrikosov, Phys Rev B **58**, 2788 (1998).
7. R. Xu, A. Husmann, T. F. Rosenbaum, M. L. Saboungi, J. E. Enderby and P. B. Littlewood, Nature **390**, 57 (1997).
8. J. S. Hu, T. F. Rosenbaum and J. B. Betts, Phys Rev Lett **95**, 186603 (2005).
9. Z. Q. Liu, W. M. Lu, X. Wang, Z. Huang, A. Annadi, S. W. Zeng, T. Venkatesan and Ariando, Phys Rev B **85**, 155114 (2012).
10. M. M. Parish and P. B. Littlewood, Nature **426**, 162 (2003).
11. G. H. Zhang, H. J. Qin, J. Chen, X. Y. He, L. Lu, Y. Q. Li and K. H. Wu, Adv Funct Mater **21**, 2351 (2011).
12. H. T. He, B. K. Li, H. C. Liu, X. Guo, Z. Y. Wang, M. H. Xie and J. N. Wang, Appl Phys Lett **100**, 032105 (2012).
13. H. Tang, D. Liang, R. L. J. Qiu and X. P. A. Gao, Acs Nano **5**, 7510 (2011).
14. D. X. Qu, Y. S. Hor, J. Xiong, R. J. Cava and N. P. Ong, Science **329**, 821 (2010).
15. X. L. Qi and S. C. Zhang, Phys Today **63**, 33 (2010).
16. M. Z. Hasan and C. L. Kane, Rev Mod Phys **82**, 3045 (2010).
17. J. E. Moore, Nature **464**, 194 (2010).
18. S. X. Zhang, L. Yan, J. Qi, M. Zhuo, Y.-Q. Wang, R. P. Prasankumar, Q. X. Jia and S. T. and Picraux, Thin Solid Films, 520, 6459 (2012).
19. A. Richardella, D. M. Zhang, J. S. Lee, A. Koser, D. W. Rench, A. L. Yeats, B. B. Buckley, D. D. Awschalom and N. Samarth, Appl Phys Lett **97**, 262104 (2010).
20. J. Chen, H. J. Qin, F. Yang, J. Liu, T. Guan, F. M. Qu, G. H. Zhang, J. R. Shi, X. C. Xie, C. L. Yang, K. H. Wu, Y. Q. Li and L. Lu, Phys Rev Lett **105**, 176602 (2010).
21. H. T. He, G. Wang, T. Zhang, I. K. Sou, G. K. L. Wong, J. N. Wang, H. Z. Lu, S. Q. Shen and F. C. Zhang, Phys Rev Lett **106**, 166805 (2011).
22. J. Wang, A. M. DaSilva, C. Z. Chang, K. He, J. K. Jain, N. Samarth, X. C. Ma, Q. K. Xue and M. H. W. Chan, Phys Rev B **83**, 245538 (2011).
23. M. H. Liu, C. Z. Chang, Z. C. Zhang, Y. Zhang, W. Ruan, K. He, L. L. Wang, X. Chen, J. F. Jia, S. C. Zhang, Q. K. Xue, X. C. Ma and Y. Y. Wang, Phys Rev B **83**, 165440 (2011).
24. H. Steinberg, J. B. Laloe, V. Fatemi, J. S. Moodera and P. Jarillo-Herrero, Phys Rev B **84**, 233101 (2011).
25. J. Chen, X. Y. He, K. H. Wu, Z. Q. Ji, L. Lu, J. R. Shi, J. H. Smet and Y. Q. Li, Phys Rev B **83**, 241304 (2011).
26. H. Z. Lu and S. Q. Shen, Phys Rev B **84**, 125138 (2011).

27. M. H. Liu, J. S. Zhang, C. Z. Chang, Z. C. Zhang, X. Feng, K. Li, K. He, L. L. Wang, X. Chen, X. Dai, Z. Fang, Q. K. Xue, X. C. Ma and Y. Y. Wang, Phys Rev Lett **108**, 036805 (2012).
28. Y. L. Chen, J. G. Analytis, J. H. Chu, Z. K. Liu, S. K. Mo, X. L. Qi, H. J. Zhang, D. H. Lu, X. Dai, Z. Fang, S. C. Zhang, I. R. Fisher, Z. Hussain and Z. X. Shen, Science **325**, 178 (2009).
29. M. von Kreutzbruck, B. Mogwitz, F. Gruhl, L. Kienle, C. Korte and J. Janek, Appl Phys Lett **86**, 072102 (2005).
30. Z. Ren, A. A. Taskin, S. Sasaki, K. Segawa and Y. Ando, Phys Rev B **82**, 241306 (2010).
31. S. Hikami, A. I. Larkin and Y. Nagaoka, Prog Theor Phys **63**, 707 (1980).
32. P. A. Lee and T. V. Ramakrishnan, Rev Mod Phys **57**, 287 (1985).
33. D. Pesin and A. H. MacDonald, Nat Mater **11**, 409-416 (2012).
34. B. A. Assaf, T. Cardinal, P. Wei, F. Katmis, J. S. Moodera and D. Heiman, arXiv:1205.4635 (2012).
35. B. F. Gao, P. Gehring, M. Burghard and K. Kern, Appl Phys Lett **100**, 212402 (2012).
36. A. D. Rata, V. Kataev, D. Khomskii and T. Hibma, Phys Rev B **68** (220403) (2003).
37. B. L. Altshuler and A. G. Aronov, Electron-electron interactions in disordered systems, Amsterdam, Netherlands : North Hollard (1985).
38. H. J. Zhang, C. X. Liu, X. L. Qi, X. Dai, Z. Fang and S. C. Zhang, Nat Phys **5**, 438-442 (2009).

**Figure Captions**

Figure 1 (a) X-ray diffraction pattern (θ-2θ scan) of a $Bi_2Te_3$ film grown on $SrTiO_3$ substrate; The inset shows the rocking curve of the (0006) peak; (b) EDX mapping of the film showing the spatial distribution of Bi (blue) and Te (red); The scale bar is 500 nm.

Figure 2 (a) The Hall resistance as a function of magnetic field; the inset shows a schematic drawing of the band structure of $Bi_2Te_3$. (b)The magnetoresistance ($MR_{xx}$) as a function of magnetic field at 1.5 and 30 K; the solid red lines are guides to the eye; The inset shows low field MR taken at different temperatures revealing the weak anti-localization effect.

Figure 3 A fit to the magneto-conductivity data at 1.5 K (a) and 30 K (b) using eq. 1 (green) and eq. 2 (blue). The insets are the zoom-in view at low fields (the regions marked by the dashed rectangles).

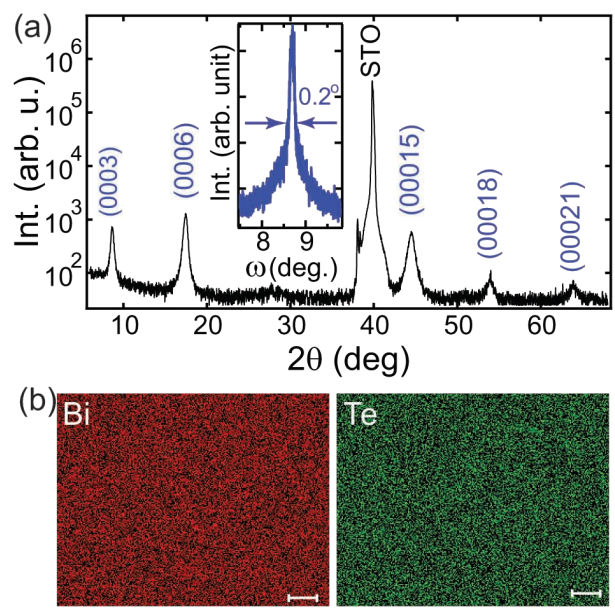

Zhang et al. Figure 1

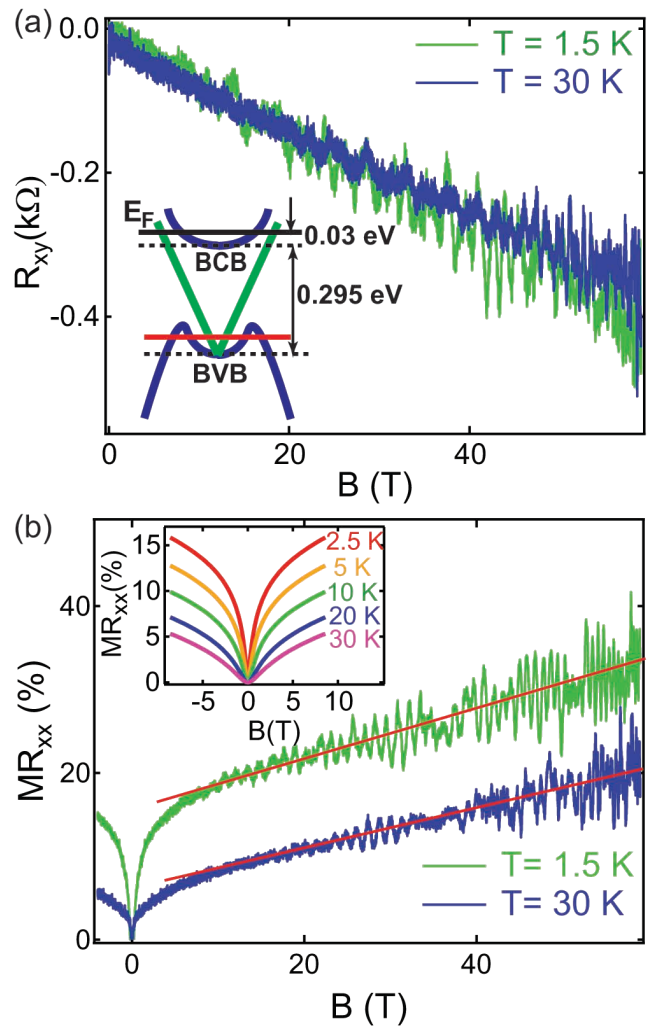

Zhang et al. Figure 2

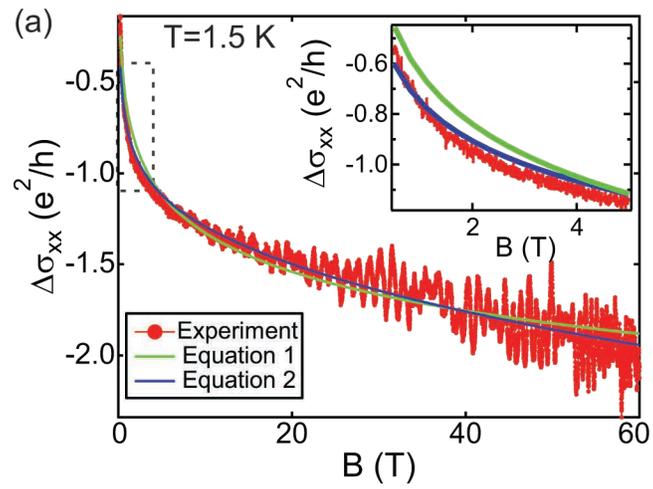

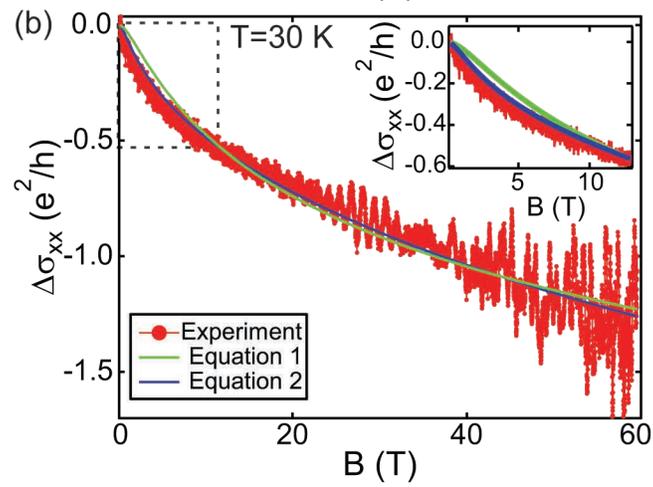

Zhang et al. Figure 3